\documentclass[cits]{PoS}

\usepackage{amsmath}
\usepackage{subfig}
\usepackage{color}

\newcommand{\msbar}{\overline{\rm MS}}

\newcommand{\mum}[1]{\ensuremath{\mu_M^{(#1)}}}
\newcommand{\mut}[1]{\ensuremath{\mu_S^{(#1)}}}

\newcommand{\nf}{{\ensuremath{N_F}}}

\vbox{
\null \vspace{0.3in}
\hbox{SMU-HEP-13-21}
}

\title{A $N_F$-Dependent VFNS for Heavy Flavors: Merging the FFNS and VFNS}

\ShortTitle{A $N_F$-Dependent VFNS for Heavy Flavors}

\author{\speaker{A. Kusina}$^a$, F.~I.~Olness$^a$, I.~Schienbein$^b$,
        T.~Je\v{z}o$^d$, K.~Kova\v{r}\'{\i}k$^c$, T.~Stavreva$^b$, J.~Y.~Yu$^b$\\
        \llap{$^a$}Southern Methodist University, Dallas, TX 75275, USA\\
        \llap{$^b$}Laboratoire de Physique Subatomique et de Cosmologie,
                   Universit\'e Joseph Fourier/CNRS-IN2P3/INPG,
                   53 Avenue des Martyrs, 38026 Grenoble, France\\
        \llap{$^c$}Institute for Theoretical Physics,
                   Karlsruhe Institute of Technology, Karlsruhe, D-76128, Germany\\
        \llap{$^d$}Department of Physics, University of Durham, Durham DH1 3LE, UK\\
                   Department of Mathematical Sciences, University of Liverpool, Liverpool L69 3BX, UK\\
        E-mail: \email{akusina@smu.edu}, \email{olness@smu.edu},
                \email{schien@lpsc.in2p3.fr}, \email{T.Jezo@liverpool.ac.uk},
                \email{kovarik@particle.uni-karlsruhe.de}, \email{stavreva@lpsc.in2p3.fr}, 
                \email{yu@physics.smu.edu}}

\abstract{
We introduce a Hybrid Variable Flavor Number Scheme (H-VFNS) for heavy
flavors,
which incorporates the advantages of both the traditional
Variable Flavor Number Scheme (VFNS) as well as the Fixed Flavor Number
Scheme (FFNS).
We include an explicit dependence on number of
active flavors $\nf$ in both the Parton Distribution Functions (PDFs)
and the strong coupling constant $\alpha_{S}$.
This results in sets of coexisting PDFs and $\alpha_{s}$ for
$\nf=\{3,4,5,6\}$, that are related analytically by the $\msbar$
matching conditions.
The H-VFNS resums the heavy
quark contributions and provides the freedom to choose the optimal
$\nf$ for each particular data set. Thus, we can fit selected HERA
data in a FFNS framework, while retaining the benefits of the VFNS
to analyze LHC data at high scales. We illustrate how such a fit can
be implemented for the case of both HERA and LHC data.
}

\FullConference{XXI International Workshop on Deep-Inelastic Scattering and Related Subjects\\
          22-26 April, 2013\\
          Marseilles, France}

\begin{document}

\section{Introduction}
\label{sec:intro}

For precision analyses of collider data, the heavy quarks (charm,
bottom, and top) must be properly taken into account; this is a
non-trivial task due to the different mass scales which enter the
theory. There are two general frameworks used for this purpose:
$(i)$ Fixed Flavor Number Scheme (FFNS),
and $(ii)$ Variable Flavor Number Scheme (VFNS).%
    \footnote{We refer here to the General Mass (GM) VFNS, where
    the mass effects are included; this is in contrast to the Zero Mass (ZM) VFNS,
    where heavy quarks are treated as massless.}

In the FFNS, 
heavy quarks are treated as extrinsic to the proton and there are no
heavy quark PDFs; they are produced only in the final state or in
loops.%
    \footnote{See refs.~\cite{Martin:2006qz, Gluck:2006ju} for
    a discussion on the details of the different formulations of the FFNS.
    Additionally an example of PDF global analysis in the FFNS can be found
    for instance in refs.~\cite{Alekhin:2009ni,Gluck:2007ck}.}
The advantage of the FFNS is the exact treatment of final state
kinematics, which is crucial near the heavy quark production
threshold. However, it is not Infra-Red (IR) safe and cannot be
extended to asymptotic scales ($\mu\gg m_Q$) because of the large
unresummed logarithms $\log(\mu/m_Q)$.

The VFNS is a set of multiple FFNS schemes with different numbers of
active flavors $N_F$, that are
connected by matching conditions.
The matching of the $N_F$ and $N_{F+1}$ schemes is performed at a matching scale
$\mu_M^{(N_F)}$, and this is traditionally  set to the heavy quark mass $m_Q$.
Thus, for $\mu$ scales  above the heavy quark mass,
a corresponding heavy quark PDF appears and resumes $\log(\mu/m_Q)$
terms; this  ensures the IR safety of the scheme.
There are multiple implementations of the VFNS  including: 
ACOT~\cite{Aivazis:1993kh,Aivazis:1993pi,Collins:1998rz,Kramer:2000hn,Tung:2001mv},
TR~\cite{Thorne:1997uu,Thorne:2006qt},
FONLL~\cite{Cacciari:1998it,Forte:2010ta},
GJR/JR~\cite{Gluck:2008gs,JimenezDelgado:2009tv};
for recent reviews see, e.g. refs.~\cite{Thorne:2008xf,Olness:2008px,Binoth:2010ra}.

In this contribution we present a new heavy flavor scheme,
denoted as the Hybrid Variable Flavor Number Scheme (H-VFNS),
which incorporates the advantages of both the traditional
VFNS as well as the FFNS.
The H-VFNS was introduced in ref.~\cite{Kusina:2013uma} and for more details
we refer reader to this reference.

\section{Hybrid Variable Flavor Number Scheme}
\label{sec:hfns}

\begin{figure*}[t]
\centering{}
\includegraphics[clip,width=0.4\textwidth]{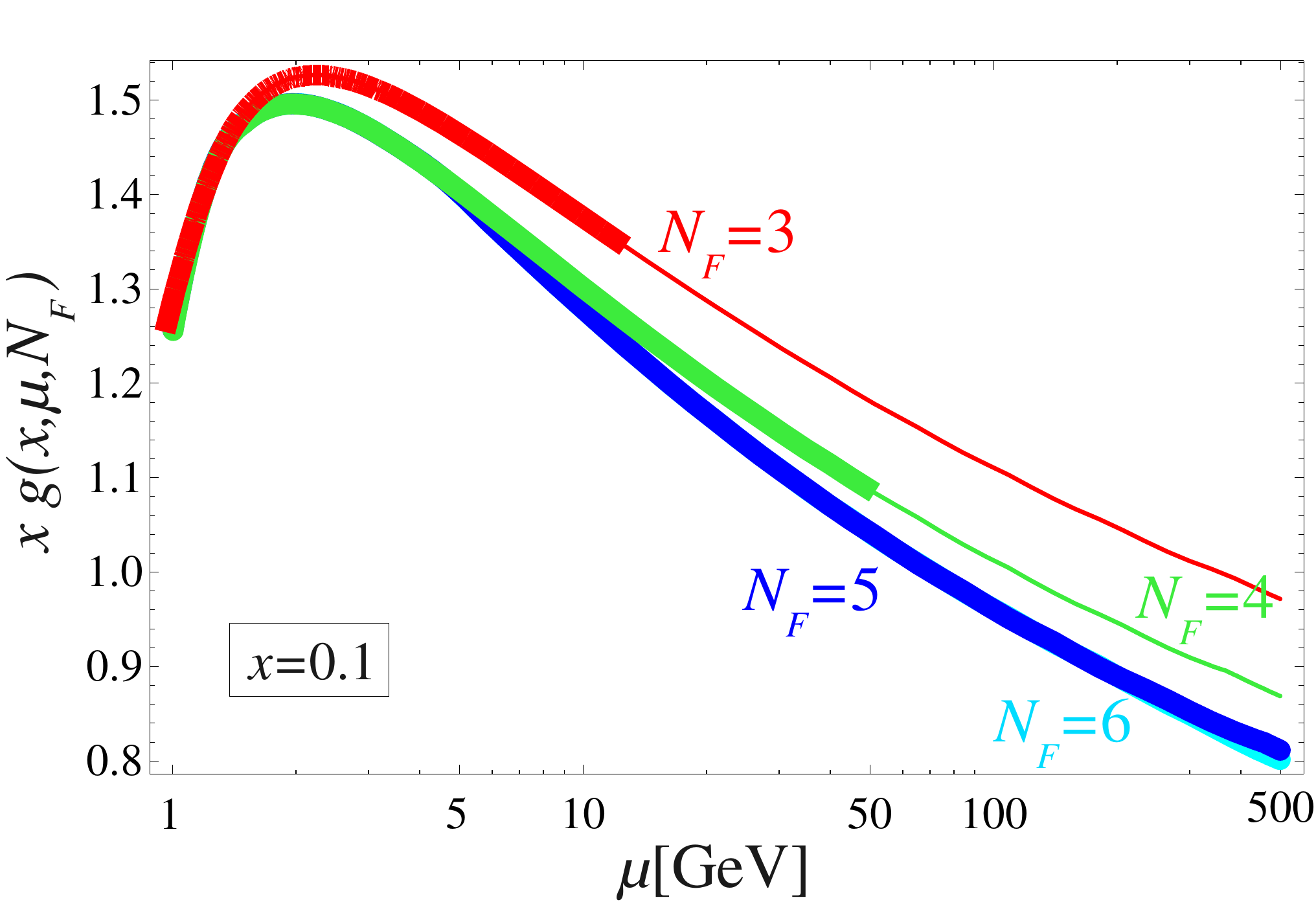}
\quad{}
\includegraphics[width=0.38\textwidth]{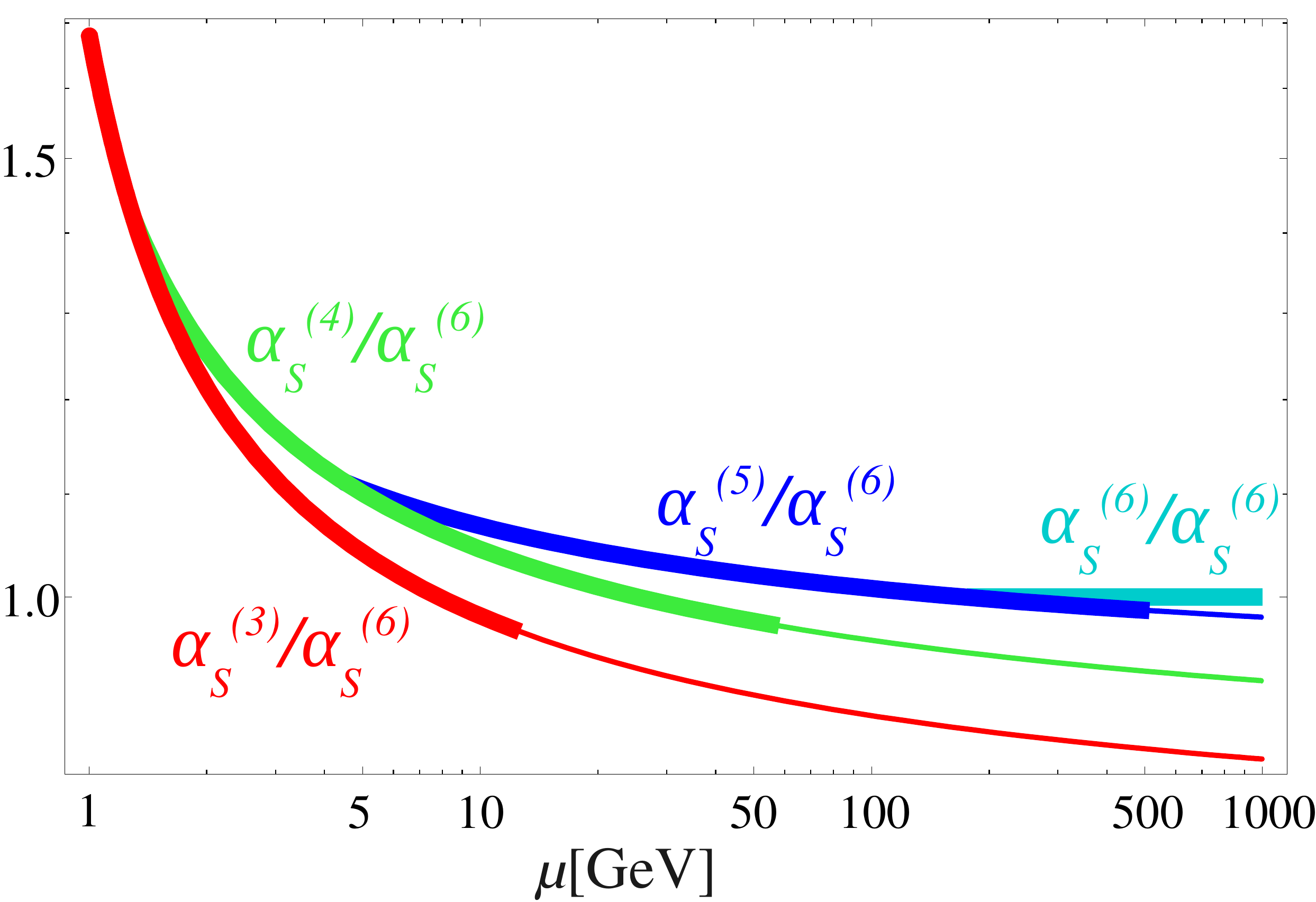}
\caption{Schematic of a H-VFNS: PDF (left) and $\alpha_{S}$ (right) vs. $\mu$.
The preferred range of each $\nf$ branch is indicated by the thicker line.}
\label{fig:schematic}
\end{figure*}

We generalize the traditional VFNS by introducing an explicit
dependence on the number of active flavors, $\nf$, in both the PDFs
$f_{a}(x,\mu,\nf)$ and the strong coupling $\alpha_{S}(\mu,\nf)$:
\begin{eqnarray*}
f_{i}(x,\mu) & \longrightarrow & f_{i}(x,\mu,\nf)\\
\alpha_{s}(\mu) & \longrightarrow & \alpha_{s}(\mu,\nf).
\end{eqnarray*}
Thus,
in the H-VFNS 
we have the freedom to choose the $\nf$ value 
at each $\mu$ scale;
this is in contrast to the traditional VFNS where the $\nf$ value is
uniquely determined by the $\mu$ scale. 
The H-VFNS is illustrated schematically in Fig.~\ref{fig:schematic}
where we explicitly see the coexistence of PDFs and $\alpha_{S}$
for different $N_{F}=\{3,4,5,6\}$ values.
%
%

On a technical level, in addition to the {\em matching scale} $\mum{\nf}$,
we introduce a separate switching scale $\mut{\nf}$.
The matching scale is a $\mu$ point where we define the $\nf+1$ PDFs and
$\alpha_{S}$ in terms of the $\nf$ ones (using the $\msbar$ matching conditions).
The \textbf{switching scale} $\mut{\nf}$ is the $\mu$ scale where we change
between $\nf$ and $\nf+1$ scheme when calculating physical observables (e.g., $d\sigma$, $F_2$).
Below the switching scale
($\mu<\mut{\nf}$) \emph{physical observables} are calculated in the
$N_F$-flavor scheme, and above the switching scale ($\mut{\nf}<\mu$)
they are calculated in the $(\nf+1)$-flavor scheme.
In contrast, in the traditional VFNS the matching and the switching
scales are equal. Indeed, in all practical applications
to date these scales have been identified with the heavy quark masses:
$\mum{\nf}=\mut{\nf}=m_{\nf}$.

The H-VFNS PDFs and $\alpha_S$ for different number of flavors are connected
analytically by the $\msbar$ matching
conditions~\cite{Buza:1996wv}.
Therefore, by knowing the PDFs for a specific $N_{F}$ branch, we
are able to compute the related PDFs for any other number of active
flavors.

Similar goal has been achieved in the frameworks of
MSTW~\cite{Martin:2006qz,Martin:2010db},
ABKM/ABM~\cite{Alekhin:2009ni}
and NNPDF~\cite{Ball:2011mu,Ball:2012cx}
by providing sets with different numbers of active flavors
that are also connected by the $\msbar$ matching conditions.
Their phenomenological implications have been recently
investigated in refs.~\cite{Thorne:2012az,Ball:2013gsa}.

\subsection{Problems resolved}
\label{subsec:why}

Since PDFs and strong couplings with different $N_F$ coexist, it
allows us to avoid dealing with a $N_{F}$ flavor transition should
it happen to lie right in the middle of a data set.
For example, if we analyze the HERA $F_{2}^{charm}$ data
(e.g.~\cite{Aktas:2005iw})
which covers a typical range of $Q\sim[3,8]$~GeV and we were to
use the traditional VFNS, then the $N_{F}$ transition between 4 and
5 flavors would lie right in the middle of the analysis region; clearly
this is very inconvenient for the analysis. Because we can specify
the number of active flavors $N_{F}$ in the H-VFNS, we have the option
to \emph{not} activate the $b$-quark in the analysis even when $\mu>m_{b}$;
instead, we perform all our calculations of $F_{2}^{charm}$ using
$N_{F}=4$ flavors. This will avoid any potential discontinuities
in the PDFs and $\alpha_{s}$ in contrast to the traditional VFNS
which forces a transition to $\nf=5$ at the $b$-quark mass.
Also since 4 and 5-flavor PDFs are connected analytically, it allows us
to use the $N_{F}=4$ PDFs extracted from the $F_{2}^{charm}$ data set
and relate this to $N_{F}=5$ PDFs that can be applied at high $\mu$ scales
for LHC processes. Note that in this example all the HERA $F_{2}^{charm}$
data (both above and below $m_{b}$) influence the $N_{F}=5$ PDFs used
for the LHC processes.

Additionally, the  H-VFNS implementation gives the user maximum flexibility
in choosing where to switch between the $N_{F}$ and $N_{F}+1$ calculations.
Not only can one choose different switching points for different processes
(as sketched above), but we can make the switching point dependent
on the kinematic variables of the process. For example, the production
thresholds for charm/bottom quarks in DIS are given in terms of the
photon-proton center of mass energy $W^{2}\simeq Q^{2}(1-x)/x$; thus,
we could use this to define our switching scales.

An important operational question is: how far above the $\mu=m_{Q}$
can we reliably extend a particular $N_{F}$ framework. We know this
will have mass singular logs of the form $\alpha_{s}\ln(\mu/m_{Q})$,
so these will eventually spoil the perturbation expansion. We just
need to ensure that we transition to the $N_{F}+1$ result before
these logs obviate the perturbation theory.
In general, we find that when the $\mu$ scale is more than a few times
the heavy quark mass then we need to be concerned with the resummation 
of these logarithms~\cite{Kusina:2013uma}.

\section{$\nf$ Dependence of the PDFs}
\label{sec:results}

One of the simplest quantities to illustrate the effect of the
number of active flavors $N_{F}$ on the PDFs $f_{i}(x,\mu,N_{F})$,
is the momentum fraction $\left[\int_{0}^{1}x\, f_{i}(x)\, dx\right]$
carried by the PDF flavors.

\begin{figure}[t]
\begin{center}
\includegraphics[width=0.4\textwidth]{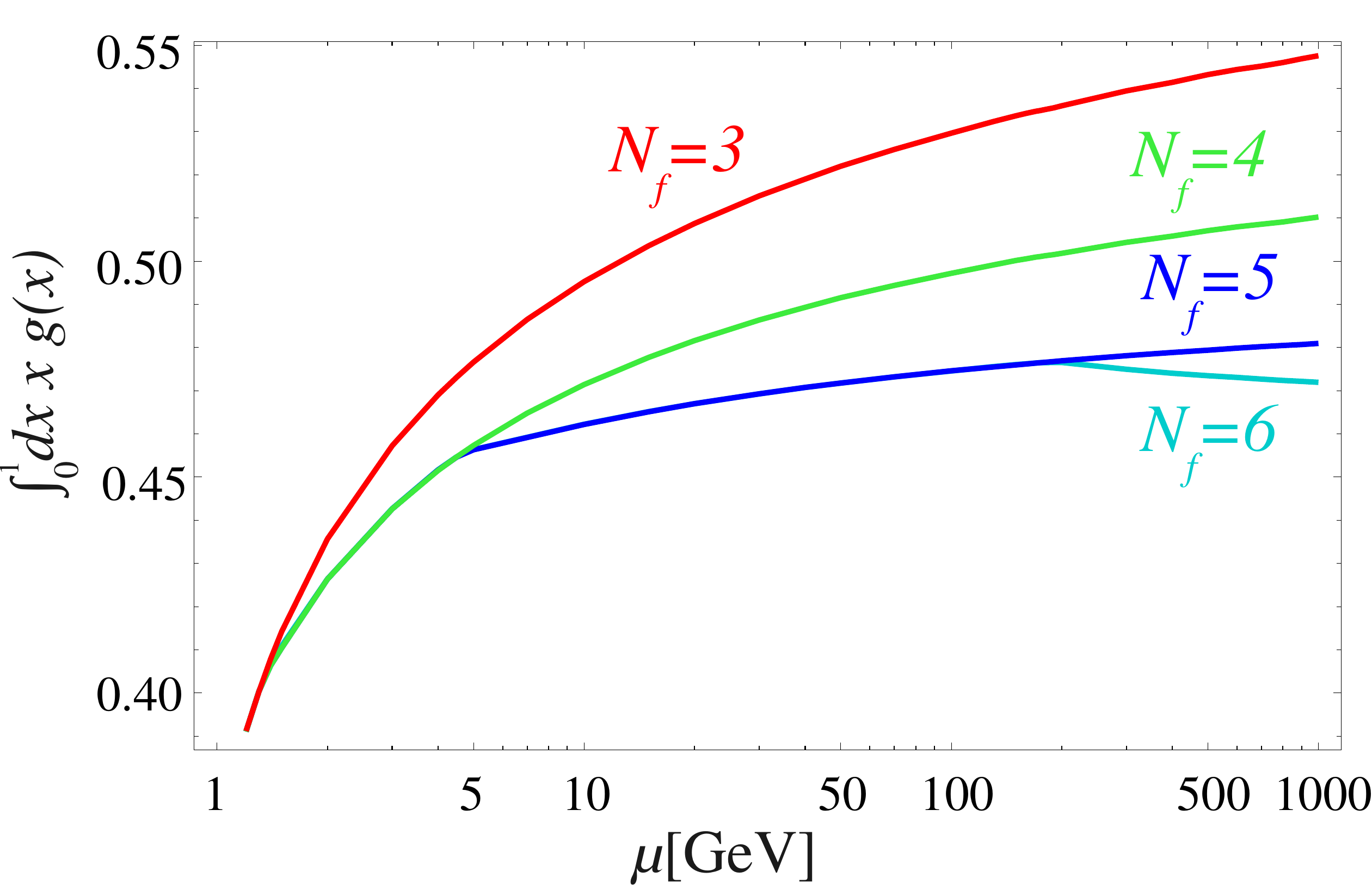}
\quad{}
\includegraphics[width=0.42\textwidth]{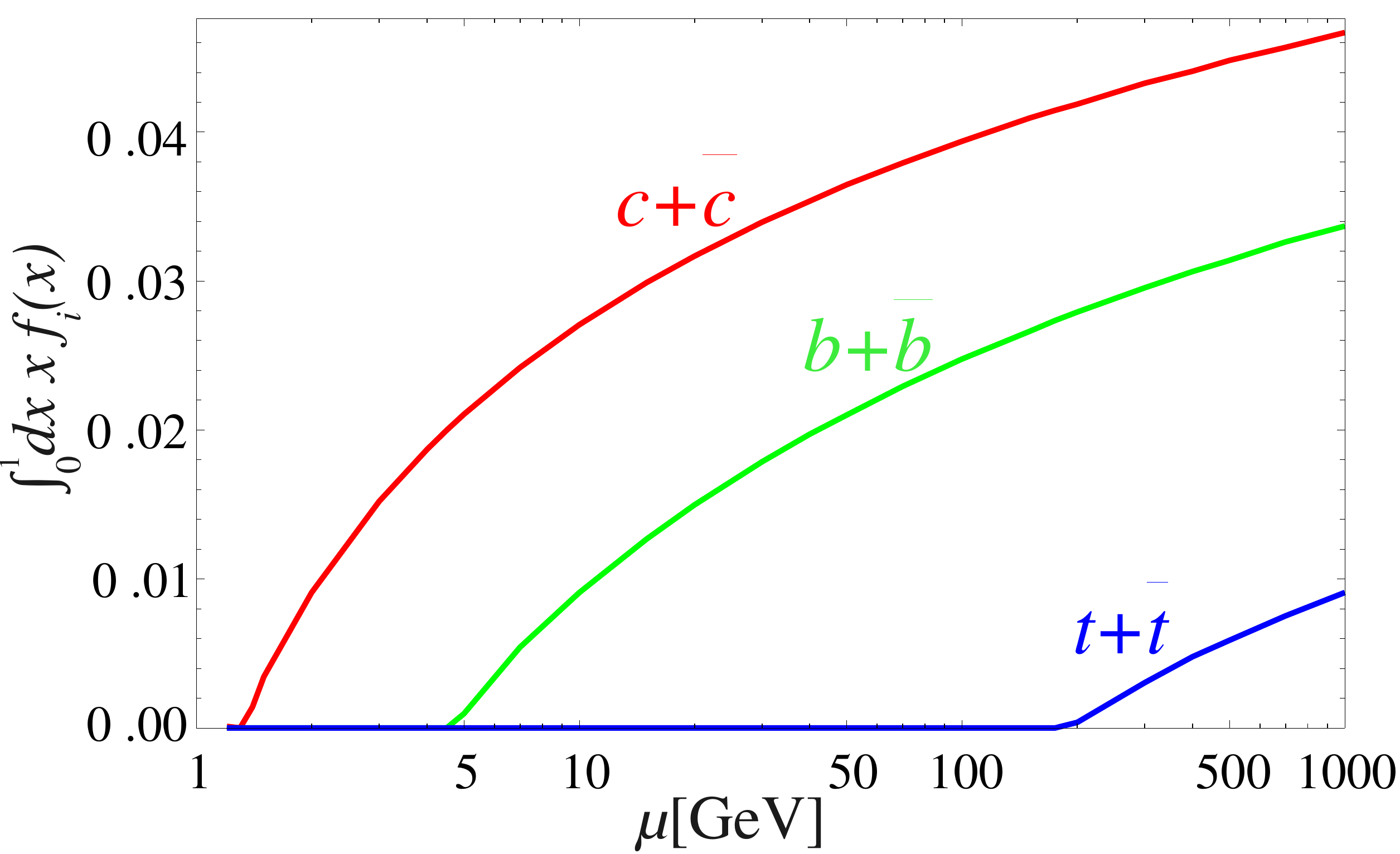} 
\end{center}
\caption{(a) Gluon momentum fraction; (b) Momentum fraction for $c+\bar{{c}}$,
$b+\bar{{b}}$ and $t+\bar{{t}}$ quarks.}
\label{fig:momFrac}
\end{figure}

Figure~\ref{fig:momFrac} shows the gluon and heavy quark momentum
fractions as a function of the $\mu$ scale. For very low $\mu$ scales
all the curves coincide by construction; when $\mu<m_{c,b,t}$ the
charm, bottom, and top degrees of freedom will ``deactivate'' and
the $N_{F}=4,5,6$ results will reduce to the $N_{F}=3$ result.
As we increase the $\mu$ scale, we open up new channels. For example,
when $\mu>m_{c}$ the charm channel activates and the DGLAP evolution
will generate a charm PDF via the $g\to c\bar{c}$ process. Because
the overall momentum sum rule must be satisfied
$\left[\sum_{i}\int_{0}^{1}x\, f_{i}(x)\, dx=1\right]$,
as we increase the momentum carried by the charm quarks, we must decrease
the momentum carried by the other partons. This interplay is evident
in Fig.~\ref{fig:momFrac}. In Figure~\ref{fig:momFrac}(a), we see
that for $\mu=1000$~GeV, the momentum fraction of the $N_{F}=4$
gluon is decreased by $\sim4\%$ as compared to the $N_{F}=3$ gluon.
Correspondingly, in Fig.~\ref{fig:momFrac}(b) we see that at $\mu=1000$~GeV,
the momentum fraction of the charm PDF is $\sim4\%$. Thus, when we
activate the charm in the DGLAP evolution, this depletes the gluon
and populates the charm PDF via $g\to c\bar{c}$ process.
The same holds for 5 and 4-flavor gluon and bottom quark.
%
%

The gluon PDF is primarily affected by the heavy $N_{F}$ channels
as it couples via the $g\to c\bar{c},b\bar{b},t\bar{t}$ processes.
The effect on the light quarks $\{u,d,s\}$ is minimal as these only
couple to the heavy quarks via higher order processes ($u\bar{u}\to g\to c\bar{c}$).

\section{An Example: From Low to High Scales}
\label{sec:example}

We now finish with an example of how the H-VFNS scheme could be employed
for a simultaneous study of both a low-scale process ($\mu\sim m_{b}$)
at HERA and a high scale process ($\mu\gg m_{c,b}$) at the LHC.

At HERA, a characteristic $Q$ range for the extraction of $F_{2}^{charm}$,
for example, is $\sim[2,10]$~GeV and this spans the kinematic region
where the charm and bottom quarks become active in the PDF. These
analyses can be performed using a $N_{F}=3$ FFNS calculation as the
scales involved are not particularly large compared to the $m_{c,b}$
scales.
Actually, the extraction of the $F_{2}^{charm}$ structure
function is often computed using the HVQDIS program~\cite{Harris:1997zq},
and this uses a $N_{F}=3$ FFNS.
Conversely, at the LHC, the $\mu$ range for the new
particle searches in the Drell-Yan process can be in excess of a TeV.
For this analysis, we would want to use
$N_{F}=5$ so that the charm and bottom logs are resummed.
Because the  H-VFNS simultaneously provides $N_{F}=\{3,4,5,6\}$,
we can analyze the HERA data in a FFNS $N_{F}=3$ context while
also analyzing the LHC data in a $N_{F}=\{4,5,6\}$ VFNS context.
Operationally, we could perform a PDF fit to both a combination of
HERA and LHC data by implementing the following steps. 
\begin{enumerate}
\item Parametrize the PDFs at a low initial scale $\mu=Q_{0}\sim1$~GeV,
and generate a family of $N_{F}$ dependent PDFs.
\item Fit the HERA $F_{2}^{charm}$ structure function data using $N_{F}=3$
``FFNS'' PDFs, $f_{i}(x,\mu,N_{F}=3)$ and $\alpha_{s}(\mu,N_{F}=3)$.
\item Fit the high-scale LHC data using $N_{F}=4,5,6$ ``VFNS'' PDFs,
$f_{i}(x,\mu,N_{F}=4,5,6)$ and $\alpha_{s}(\mu,N_{F}=4,5,6)$. 
\item Repeat steps 1) through 3) until we have a suitable minimum.
\end{enumerate}
Note, because we generate all the PDFs and $\alpha_{s}$ for all $N_{F}=\{3,4,5,6\}$
flavors in step 1), the separate $N_{F}$ branches are analytically
related.
Also because we have access to all $N_{F}=\{3,4,5,6\}$
sets, there is no difficulty in performing the HERA analysis of step
2) and the LHC analysis of step 3) in different $N_{F}$ frameworks.

Finally,
the user is now responsible for ensuring each $N_{F}$ calculation
is {\em not} used beyond its range of validity. While it is now possible to
compute with $N_{F}=3$ at high $\mu$ scales, this is not necessarily
a reliable result.

\subsection{$N_{F}$ Conversion Factors}

Finally, we demonstrate how to use the family of $N_{F}$ dependent
PDFs to estimate the effect of changing from $N_{F}=3$ to $N_{F}=5$
in a calculation such as the extraction of $F_{2}^{charm}$ discussed
above.
For example, the HVQDIS program~\cite{Harris:1997zq}
uses a $N_{F}=3$ FFNS while many of the PDFs are only available
for $N_{F}=4,5$. If we have access to both
$N_{F}=3$ and $N_{F}=5$ PDFs, we can simply use the correct $N_{F}$
PDF set, and the conversion between the different $N_{F}$ sets is
simply given by the following identity: 
$
f^{(N_{F}=5)}(x) = f^{(N_{F}=3)}(x)
[
f^{(N_{F}=5)}(x)
/
f^{(N_{F}=3)}(x)
].
$
The term in brackets above represents the ``correction factor''
in converting between $N_{F}=3$ and $N_{F}=5$ PDF sets.
As we noted in Sec.~\ref{sec:results}, the dominant effect of changing
from $N_{F}=3$ to $N_{F}=5$ was to deplete the gluon PDF which fed
the charm PDF via the $g\to c\bar{c}$ process. Therefore, we can
estimate this effect by comparing the shift of the gluon PDF for $N_{F}=3$
and $N_{F}=5$.
So even if we do not have access to both the $N_{F}=3$ and $N_{F}=5$
PDF sets, the combination $[f^{(N_{F}=5)}/f^{(N_{F}=3)}]$ is driven
by the DGLAP evolution and only mildly sensitive to the detailed PDF;
hence, the above technique can still provide a rough approximation
as to the correction factor between the $N_{F}=3$ and $N_{F}=5$
PDFs.

\section{Conclusion}
\label{sec:conclusions}

We have investigated the $N_{F}$ dependence of the
PDFs and proposed an extension of the traditional VFNS which we denote
the H-VFNS. In this scheme, we include an explicit $N_{F}$ dependence
in both the PDFs $f_{a}(x,\mu,N_{F})$ and strong coupling $\alpha_{S}(\mu,N_{F})$;
this provides the user the freedom, and responsibility, to choose
the appropriate $N_{F}$ values for each data set and kinematic region.
Thus, the H-VFNS provides a valuable tool for fitting data across a wide
variety of processes and energy scales from low to high.

\section*{Acknowledgments}
We thank
Sergey Alekhin,
Michiel Botje,
John Collins,
Kateria Lipka,
Pavel Nadolsky,
Voica Radescu,
Randall Scalise, 
and
the members of the HERA-Fitter group
for valuable discussions.
F.I.O., I.S., and J.Y.Y. acknowledge the hospitality of CERN, DESY,
Fermilab, and Les Houches where a portion of this work was performed.
This work was partially supported by the U.S. Department of Energy
under grant {DE-FG02-13ER41996}, and the Lighter Sams
Foundation. The research of T.S. is supported by a fellowship from
the Th\'eorie LHC France initiative funded by the CNRS/IN2P3. This work
has been supported by \textit{Projet international de cooperation
scientifique} PICS05854 between France and the USA. T.J. was supported
by the Research Executive Agency (REA) of the European Union under
the Grant Agreement number PITN-GA-2010-264564 (LHCPhenoNet).

\bibliographystyle{utphys_spires}
\bibliography{bibNF}

\end{document}